\documentclass[aps, prc, onecolumn, 
showpacs, showkeys, 
preprint, preprintnumbers, 
groupedaddress,
nofootinbib
]{revtex4-1}

\usepackage[utf8]{inputenc}
\usepackage[T1]{fontenc}
\usepackage{lmodern}
\usepackage{microtype}
\usepackage[english]{babel}

\usepackage{amssymb,amsbsy,amsmath,amsfonts,amsthm}

\usepackage{yhmath} 

\usepackage{slashed}
\usepackage{bm}
\usepackage{times}

\usepackage[dvipsnames,table,xcdraw]{xcolor}
\definecolor{lightyellow}{RGB}{255,250,205}

\usepackage{graphicx}
\usepackage{epstopdf}
\usepackage{tabularx}
\usepackage{multirow}
\usepackage{longtable}

\usepackage{hyperref}
\hypersetup{
  colorlinks=true,
  linkcolor=blue,  
  citecolor=cyan,
  urlcolor=red,           
}

\usepackage{url}

\def\orcid#1{\kern .08em\href{https://orcid.org/#1}{\includegraphics[keepaspectratio,width=0.7em]{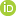}}}

\begin{document}

\title{Nucleon-nucleon scattering up to next-to-next-to-leading order in manifestly Lorentz-invariant chiral effective field theory: peripheral phases}

\author{Xiu-Lei Ren\orcid{0000-0002-5138-7415}}
\affiliation{Institut f\"ur Kernphysik \&  Cluster of Excellence PRISMA$^+$,
 Johannes Gutenberg-Universit\"at  Mainz,  D-55128 Mainz, Germany}
 \affiliation{Helmholtz Institut Mainz, D-55099 Mainz, Germany}

\author{E.~Epelbaum
\orcid{0000-0002-7613-0210}
}
 \affiliation{Institut f\"ur Theoretische Physik II, Ruhr-Universit\"at Bochum,  D-44780 Bochum,
 Germany}
\author{J.~Gegelia
\orcid{0000-0002-5720-9978}
}
 \affiliation{Institut f\"ur Theoretische Physik II, Ruhr-Universit\"at Bochum,  D-44780 Bochum,
 Germany}
\affiliation{High Energy Physics Institute, Tbilisi State  University,  0186 Tbilisi,
 Georgia}

\begin{abstract}
    We study the nucleon-nucleon interaction up to
    next-to-next-to-leading order using time-ordered perturbation
    theory in the framework of manifestly Lorentz-invariant chiral effective field theory. 
  We present the two-pion exchange contribution at one-loop level,
  which is consistent with the corresponding non-relativistic
  expressions in the large-nucleon-mass limit. Using the Born series truncated at one-loop order, 
    we calculate the phase shifts and mixing angles of the partial
    waves with the angular momentum $l\geq 2$. 
Comparing with the results of non-relativistic formulation, we find an
improved description of the phase shifts for some $D$ waves  such as 
the $^3D_3$ one. 
For the other partial waves, both approaches show the  globally similar results.
      
\end{abstract}


\maketitle

\date{\today}

\section{Introduction}

Chiral effective field theory (ChEFT) for few-nucleon systems goes back to  the
seminal papers by Steven Weinberg, who extended chiral
perturbation theory~\cite{Weinberg:1978kz} to systems involving two and more nucleons
\cite{Weinberg:1990rz,Weinberg:1991um}.
In the resulting ChEFT approach,  
the power counting rules
are applied to the effective potentials defined
as sums of contributions of few-nucleon-irreducible diagrams.
The scattering amplitudes  are then obtained by solving the
Schr\"odinger equation or the corresponding integral equation in
momentum space. 
For reviews of ChEFT in the few-body sector see
Refs.~\cite{Bedaque:2002mn,Epelbaum:2005pn,Epelbaum:2008ga,Machleidt:2011zz,Epelbaum:2012vx,Hammer:2019poc}. 

In ChEFT one can only reliably calculate the effective potential for small momenta corresponding  to its long-range part in the coordinate space \cite{Weinberg:1990rz,Weinberg:1991um}.
Loop integrals of scattering equations, on the other hand, involve  integration over all momenta including the ultraviolet (UV) region. Naive extensions of the long-range parts of the chiral effective potentials to short distances 
result in singular potentials causing severe problems when solving
equations. Such singular potentials generate deeply bound states that
are absent in the underlying theory \cite{Landau:1991wop}. 
The singular behavior of the chiral potentials at short distances
therefore clearly represents an artifact of a naive extrapolation of the long-range potential to short distances \cite{Epelbaum:2018zli}. 
While the resulting arbitrariness should not influence physical
observables if the renormalization is carried out properly  by an appropriate treatment of the short range components 
encoded in contact interactions of the effective Lagrangian,
in practice, carrying out correct quantum field theoretical
renormalization in the few-body sector of ChEFT turned out to be a
challenging problem. 
Conflicting points of view about  this issue do not seem to converge
to a consensus even after two decades of intense research. 
We refer interesting readers to Refs.~\cite{Weinberg:1991um,Gegelia:2004pz,Epelbaum:2006pt,Epelbaum:2009sd,Epelbaum:2017byx,Epelbaum:2018zli,Epelbaum:2020maf,Valderrama:2016koj,Hammer:2019poc,vanKolck:2020llt,Epelbaum:2021sns,Gasparyan:2021edy,
Lepage:1997cs,Gegelia:1998iu,Park:1998cu,Lepage:1999kt,Epelbaum:2004fk,Gegelia:2004pz,Kaplan:1996xu,Beane:1997pk,Kaplan:1998tg,Birse:1998dk,
Nogga:2005hy,Long:2007vp,Birse:2009my,Harada:2010ba,
Long:2011qx,Long:2011xw,Harada:2013hga,Beane:2021dab,Tews:2022yfb,Frederico:1999ps,Timoteo:2005ia,Yang:2007hb,Yang:2009kx,Beane:2000wh,Wang:2020myr,Valderrama:2009ei,Long:2012ve,vanKolck:2020plz,vanKolck:2021rqu,Griesshammer:2021zzz}
for a collection of different points of view on this issue. 

Aiming at an improved ultraviolet behavior of the effective potential,
i.e.~at its different UV extension, while keeping the same infrared
(IR) behavior  as in the standard non-relativistic formalism,  
a modified  Weinberg's approach to the
nucleon-nucleon (NN) scattering problem has been proposed in
Ref.~\cite{Epelbaum:2012ua}. This novel scheme employs time-ordered
perturbation theory (TOPT) and
relies on the manifestly Lorentz-invariant effective Lagrangian. It is
important to emphasize that it was not thought as a replacement of the
non-relativistic formalism but rather, being equivalent  
in the IR region, it results in a less divergent UV behavior leading to a perturbatively renormalizable modification of Weinberg's approach.
Based on this idea, a systematic framework for chiral nuclear forces with detailed diagrammatic rules of TOPT 
for particles with non-zero spin and interactions involving time derivatives has been worked out in Ref.~\cite{Baru:2019ndr}.
These rules can be applied systematically at all orders in the loop expansion. Analogously to the non-relativistic approach,   
using  the standard Weinberg
power counting for diagrams contributing to NN
scattering one has to take into account an infinite number of graphs
already at leading order (LO). The infinite series of diagrams can be
resummed by defining the effective potential as a sum of all
two-nucleon-irreducible TOPT diagrams and substituting into the
Kadyshevsky integral equation \cite{kadyshevsky}, which results directly from the TOPT rules. 
The scattering amplitude can be calculated order-by-order using both a
renormalizable approach, which relies on a perturbative treatment of
corrections beyond LO and allows one to completely eliminate the
UV cutoff, and a conventional scheme based on iterating a truncated
potential to all orders while keeping the UV cutoff parameter of the order of the hard scale of the problem
\cite{Lepage:1997cs,Gegelia:1998iu,Park:1998cu,Lepage:1999kt,Epelbaum:2004fk,Gegelia:2004pz,Epelbaum:2006pt,Epelbaum:2018zli}.
In the latter case, the formulation based on the manifestly
Lorentz-invariant effective Lagrangian is expected to permit a larger cutoff variation as compared
to the non-relativistic approach thanks to the improved UV behavior. 

To explicitly verify the above expectations  it is necessary to go beyond LO in our manifestly Lorentz-invariant formulation of Ref.~\cite{Baru:2019ndr}.\footnote{Different paths of calculating NN potential up to NNLO using relativistic ChEFT has 
been taken in Refs.~\cite{Higa:2003jk, Higa:2003sz,
  Ren:2016jna,Xiao:2020ozd,Wang:2021kos,Lu:2021gsb}.} Therefore, the
main purpose of this paper is to derive the two-pion exchange (TPE)
contributions to NN potential
up to next-to-next-to-leading order (NNLO), which are expected to
describe the medium-range part of the NN interaction. 
  Notice here that due to the absence of chiral-order-one ($\nu=1$)
  contributions to the NN potential, the term NNLO refers to chiral
  order three ($\nu=3$). In analogy to Ref.~\cite{Kaiser:1997mw},
  taking into account the fact that for higher partial waves the NN
  potential becomes weaker, we assume that a perturbative treatment should be adequate in this case. Using the Born series truncated at one-loop order 
we calculate the phase shifts and mixing angles of the partial waves
with orbital angular momentum $l\geq 2$ and compare our results with the empirical phase shifts as well as to the results of the non-relativistic ChEFT.

Our paper is organized as follows: in section~\ref{Section2} we specify the effective Lagrangian, give the diagrammatic rules of TOPT,  and 
work out  the details of the NN potential. Various checks of the obtained effective potential are performed and the phase shifts of peripheral 
partial waves are calculated and compared to analogous 
results in the non-relativistic formalism    in section~\ref{Section3}. 
The results of our work are summarized in  section~\ref{Summary}.

\section{Two-pion-exchange contributions}
\label{Section2}
In this section, we present the TPE contributions to the NN scattering amplitude at one-loop order. 
We start with the Lorentz-invariant effective chiral Lagrangian and briefly summarize the corresponding diagrammatic rules of TOPT  obtained in Ref.~\cite{Baru:2019ndr}. 
Using the Weinberg power counting we identify all TPE diagrams contributing to the scattering amplitude up to NNLO. Ultraviolet divergences and power-counting 
violating pieces of loop diagrams are removed by using the subtractive renormalization.

\subsection{Effective chiral Lagrangian}
The Lorentz-invariant effective chiral Lagrangian required for calculating one-loop contributions to the NN potential up to NNLO is given by
\begin{equation}
	\mathcal{L}_\mathrm{eff} = \mathcal{L}_{\pi\pi}^{(2)} + \mathcal{L}_{\pi N}^{(1)} + \mathcal{L}_{\pi N}^{(2)},
\end{equation}
where the superscripts denote the chiral orders, and
\cite{Gasser:1984yg,Bellucci:1994eb,Gasser:1987rb,Fettes:2000gb}
\begin{equation}
\begin{aligned}
\mathcal{L}_{\pi\pi}^{(2)} &= \frac{f_\pi^2}{4}\langle u_\mu u^\mu + \chi_+\rangle, \\
\mathcal{L}_{\pi N}^{(1)} &=\bar{\Psi}_{N}\left\{i \slashed D -m_N+\frac{g_A}{2}\, \slashed u\, \gamma^{5}\right\} \Psi_{N}, \\
\mathcal{L}_{\pi N}^{(2)} &=\bar{\Psi}_{N}\left\{c_{1}\langle\chi_+\rangle-\frac{c_{2}}{4 m_N^{2}}\left\langle u^{\mu} u^{\nu}\right\rangle\left(D_{\mu} D_{\nu}+\text { h.c. }\right)+\frac{c_{3}}{2}\left\langle u^{\mu} u_{\mu}\right\rangle-\frac{c_{4}}{4} \gamma^{\mu} \gamma^{\nu}\left[u_{\mu}, u_{\nu}\right]\right\} \Psi_{N}\,.
\end{aligned}
\end{equation} 
Leaving out the external sources, we have $u_\mu=i(u^\dagger \partial_\mu u - u \partial_\mu u^\dagger)$ and $\chi_+=u^\dagger \chi u + u\chi u^\dagger$, 
with $u=\mathrm{exp}(i\Phi/2 f_\pi)$ and $\chi=\mathrm{diag}(M_\pi^2, M_\pi^2)$. The chiral covariant derivative acting on $\Psi_N$ is given by $D_\mu \Psi_N = \partial_\mu\Psi_N + [\Gamma_\mu, \Psi_N]$ 
with $\Gamma_\mu = \frac{1}{2}\left(u^\dagger\partial_\mu u + u\partial_\mu u^\dagger\right)$. The pion and nucleon fields are collected in
\begin{equation}
  \Phi = \left(\begin{array}{cc}
           \pi^0 & \sqrt{2}\pi^+ \\
             \sqrt{2}\pi^- & -\pi^0
          \end{array}
  \right), \quad    \Psi_N = \left(
                 \begin{array}{c}
                   p \\
                   n \\
                 \end{array}
               \right).
\end{equation}
The above specified effective Lagrangian depends on the following parameters: the pion decay constant~$f_\pi$, the axial vector coupling $g_A$, 
the pion mass $M_\pi$, the nucleon mass  $m_N$ and the four low-energy constants (LECs) $c_{1}$,$c_2$, $c_3$, $c_4$, introduced in the $\pi N$ Lagrangian 
of the second order,  $\mathcal{L}_{\pi N}^{(2)}$. Numerical values of these parameters will be specified in the next section.

 
\subsection{Diagrammatic rules of TOPT}
To derive the NNLO chiral potential, we apply the diagrammatic rules
of TOPT to the manifestly Lorentz-invariant effective Lagrangian. These rules were 
obtained in Ref.~\cite{Baru:2019ndr} and  are briefly summarized  below. 

For the elastic NN scattering process, $N(p_1)+N(p_2)\to N(p_3) + N(p_4)$, the $S$ matrix can be written as 
\begin{equation}
 S= 1 - (2\pi)^4\, i\, T\, \delta^{4}(p_3+p_4-p_1-p_2) \, \prod_{i=1}^4 (2\pi)^{-3/2} \left(\frac{m_N}{\omega_{p_i}}\right)^{1/2}, 	
\end{equation}
where the four-momenta of the initial and final states are $p_1^\mu=(\omega_p, \bm{p})$, $p_2^\mu=(\omega_p, -\bm{p})$, $p_3^\mu=(\omega_{p'}, \bm{p}')$, 
and $p_4^\mu=(\omega_{p'}, -\bm{p}')$. Here $\bm{p}$ and $\bm{p'}$ are the three-momenta of the incoming and outgoing nucleons in the center-of-mass frame, respectively, and the nucleon energy is defined as 
$\omega_l : = \sqrt{\bm{l}^2+m_N^2}$.
The (on-shell) scattering amplitude $T$ can be given as a sum of an infinite number of time-ordered diagrams. Contribution of each diagram is evaluated via the following diagrammatic rules:  
\begin{itemize}
\item Draw all possible time-ordered diagrams contributing to NN
  scattering at a given order in the loop expansion and having the same combination of coupling constants;
\item Assign to each incoming (outgoing) external nucleon line with momentum $p_i$ ($p_i'$) Dirac spinor $u(p_i)$ ($\bar{u}(p_i')$);
\item Assign to each internal nucleon line a factor
\begin{equation}
  \frac{m_N}{\omega_{p_i}}\sum u(p_i)\bar{u}(p_i),	
\end{equation}
where the sum is carried out over polarizations; 
\item Assign to each internal anti-nucleon-line a factor 
 \begin{equation}
  \frac{m_N}{\omega_{p_i}}\sum u(p_i)\bar{u}(p_i)-\gamma_0,	
\end{equation}
where $\gamma_0$ is the  Dirac gamma  matrix;
\item Each internal pion line gives a factor  
\begin{equation}\frac{1}{2\varepsilon_{p_\pi}}	
 \end{equation}
 with the pion
 four-momentum $p_\pi$ and the pion energy defined as $\varepsilon_l := \sqrt{\bm{l}^2 + M_\pi^2}$;	
\item Each intermediate state gives an energy denominator  
\begin{equation}
	\frac{1}{E-\sum\limits_i \omega_{p_i}-\sum\limits_j \varepsilon_{p_j} + i\epsilon } ,
\end{equation} 
where $E$ is the total energy of the NN system and the indices $i$,
$j$ label the internal nucleon/pion lines in the intermediate state;
\item Each interaction vertex is obtained using the standard Feynman rules, special care needs to be taken of the zeroth components of momenta appearing in vertices. 
Details can be found in Ref.~\cite{Baru:2019ndr};
\item While each one-loop diagram with internal momentum $k$  contains a three-dimensional integration 
\begin{equation}
 	\int \frac{d^3 k}{(2\pi)^3}\,.
 \end{equation}

\end{itemize}

\subsection{The NN $T$-matrix at one loop  order}
Here we present the results for the  NN $T$-matrix at one-loop order. It contains two pieces: the TPE potential  and the once-iterated one-pion-exchange (OPE) potential. We use this expression of the amplitude for calculating
  phase shifts of higher partial waves. If we assume that the
  potential in 
peripheral
  partial waves 
is suppressed, due to the centrifugal barrier, by two chiral orders compared to $S$ and $P$ waves, then 
the TPE potential and the once-iterated OPE potential are of the same order and further iterations of the OPE potential are of higher chiral orders. Notice that such an assumption is compatible with Weinberg power counting which assigns 
specific chiral orders to the potential expressed in the plane wave
basis, i.e.~to the sum of all partial waves. Furthermore, the suppression of OPE within chiral EFT has been intensively studied in Refs.~\cite{Birse:2005um,PavonValderrama:2016lqn,Wu:2018lai,Kaplan:2019znu}. 
Notice that the one-loop corrections to the OPE potential 
are included by
expressing the OPE potential in terms of physical coupling constants.

  \begin{figure}[t]
\includegraphics[width=0.6\textwidth]{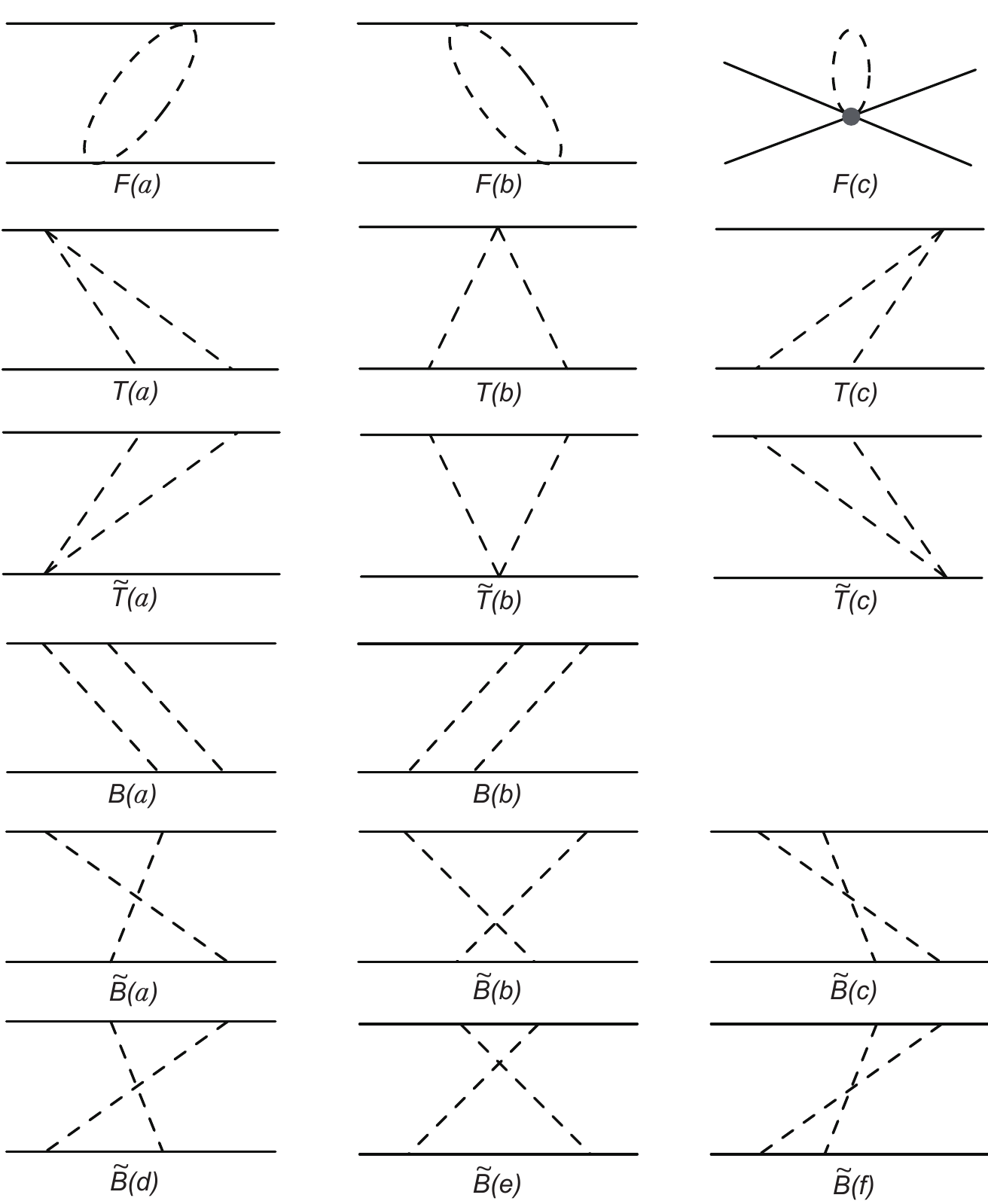}
  \caption{Time-ordered diagrams contributing to the TPE potential at NLO. Solid and dashed lines correspond to nucleons and pions, respectively. The pion-nucleon vertices are from $\mathcal{L}_{\pi N}^{(1)}$.}
  \label{Fig:LO_TPE}
\end{figure}

\subsubsection{Two-pion exchange potential}

We apply the standard Weinberg power counting to derive the chiral potential~\cite{Weinberg:1990rz,Weinberg:1991um}. 
The TPE potential at one-loop order contains two parts: $V_{2\pi}^{(\nu=2)}$ and $V_{2\pi}^{(\nu=3)}$,
where $\nu$ indicates chiral orders of time-ordered diagrams, 
\begin{equation}
\nu=2 l+\sum_{i}V_i \left(d_{i}+\frac{n_{i}}{2}-2\right),
\end{equation}
where $l$ is the number of loops, $V_i$ is the number of vertices of
type $i$, $d_{i}$ is the number of
derivatives acting on pion fields and/or spatial components of derivatives acting on nucleon fields, or pion-mass insertions, and $n_{i}$
denotes the number of nucleon fields involved in vertex $i$. The sum
in the above equation runs over all vertices contained in the diagram.

At second order ($\nu=2$), the two-nucleon irreducible time-ordered diagrams contributing to the TPE potential are shown in Fig.~\ref{Fig:LO_TPE},
\begin{equation}
	V_{2\pi}^{(\nu=2)} = V_F^{(2)} + V_{T+\tilde{T}}^{(2)} + V_B^{(2)} + V_{\tilde{B}}^{(2)}.
\end{equation}
 Using the TOPT rules and keeping only the LO terms in the expansion
 of the Dirac spinors in small momenta,   
\begin{equation}
  u = u_0 + u_1 + \cdots , \quad \bar{u}=\bar{u}_0+\bar{u}_1 + \cdots ,	
\end{equation}
where $u_0=P_+ \, u(p)$ and $\bar{u}_0=\bar{u}(p)\, P_+$  with $P_+\equiv (1+\slashed v)/2$ and $v=(1,0,0,0)$, 
 we obtain the following expressions:  
\begin{itemize}
\item Contribution of the football diagrams [the sum of diagrams $F$(a)-$F$(c) in Fig.~\ref{Fig:LO_TPE}] 	 
  \begin{equation}
  V_F^{(2)} = \frac{\bm{\tau}_1\cdot\bm{\tau}_2 }{16f_\pi^4}\int \frac{d^3k}{(2\pi)^3} \frac{(\varepsilon_k+\varepsilon_{k+q})(\omega_p+\omega_{p'})+4\varepsilon_k\varepsilon_{k+q}-E(\varepsilon_k+\varepsilon_{k+q})}{2 \varepsilon_{k} \varepsilon_{k+q}\left(\varepsilon_{k}+\varepsilon_{k+q}+\omega_p +\omega_{p'}-E\right)},
\end{equation}
where $\bm{\tau}_i$ denote the isospin Pauli matrices of nucleon $i$, $E$ is the total energy of the two-nucleon system, and
the momentum transfer is given by $\bm{q}=\bm{p}'-\bm{p}$.
Notice that the third diagram, $F(c)$, is due to the zeroth component
of momentum appearing in the vertex corresponding to the
Weinberg-Tomozawa interaction \cite{Weinberg:1991um,Epelbaum:2007us}.

\item Contribution of triangle diagrams [the sum of diagrams $T (a)-T(c)$ and $\tilde{T}(a)-\tilde{T}(c)$ in Fig.~\ref{Fig:LO_TPE}] 	
\begin{eqnarray}
	V_{T+\tilde{T}}^{(2)}
&=& \frac{4 m_N g_A^2\bm{\tau}_1\cdot\bm{\tau}_2}{128 f_{\pi}^{4}} 
\int \frac{d^3 k}{(2\pi)^3}\left[ 
\left(\bm{k}^{2}+(\bm{p}^{\prime}-\bm{p}) \cdot \bm{k}\right) + \frac{i(\bm{\sigma}_1+\bm{\sigma}_2)\cdot \bm{n}}{2}\,(a+b) \right] \\
& \times& 
\left[ \frac{\varepsilon_{k+q}-\varepsilon_{k}}{\varepsilon_k \varepsilon_{k+q} \omega_{p-k}}\left(\frac{1}{D_{T_{a}}} + \frac{1}{D_{\tilde{T}_{a}}}  - \frac{1}{D_{T_{c}}} - \frac{1}{D_{\tilde{T}_{c}}} \right) 
+  \frac{\varepsilon_{k}+\varepsilon_{k+q}}{\varepsilon_k \varepsilon_{k+q} \omega_{p-k}}
\left(\frac{1}{D_{T_{b}}} + \frac{1}{D_{\tilde{T}_{b}}} \right)
\right],
\nonumber
\label{trD}
\end{eqnarray}
where $\bm{\sigma}_i$ refer to the spin Pauli matrices of the nucleon
$i$, $\bm{n}=\bm{p}\times\bm{p}'$, and the denominators corresponding to intermediate states are given by:  
\begin{equation}
\begin{aligned}
D_{T_a} &= \left(E-\omega_{k-p}-\varepsilon_{k+q}-\omega_{p'}\right)\left(E-\varepsilon_{k+q}-\varepsilon_{k}-\omega_{p}-\omega_{p'}\right), \\ 
D_{T_b} &= \left(E-\omega_{k-p}-\varepsilon_{k}-\omega_{p}\right)\,\left(E-\omega_{k-p}-\varepsilon_{k+q}-\omega_{p'}\right), \\
D_{T_c} &= \left(E-\omega_{k-p}-\varepsilon_{k}-\omega_{p}\right)\,\left(E-\varepsilon_{k+q}-\varepsilon_{k}-\omega_{p}-\omega_{p'}\right), \\ 
D_{\tilde{T}_a} &= \left(E-\omega_{k-p}-\varepsilon_{k+q}-\omega_{p'}\right)\,\left(E-\varepsilon_{k+q}-\varepsilon_{k}-\omega_{p}-\omega_{p'}\right)	,\\
D_{\tilde{T}_b} &= \left(E-\omega_{k-p}-\varepsilon_{k}-\omega_{p}\right)\, \left(E-\omega_{k-p}-\varepsilon_{k+q}-\omega_{p'}\right) ,\\ 
D_{\tilde{T}_c} &= \left(E-\omega_{k-p}-\varepsilon_{k}-\omega_{p}\right)\,\left(E-\varepsilon_{k+q}-\varepsilon_{k}-\omega_{p}-\omega_{p'}\right)\,.
\end{aligned}
\end{equation}
The parameters $a$ and $b$ in Eq.~(\ref{trD}) stand for the coefficients of the decomposition $\bm{k}= a\,\bm{p} + b\, \bm{p}' + c\,  (\bm{p}'\times \bm{p})$, where
\begin{equation}
a=\frac{\bm{p}^{\prime}\cdot\bm{p} \, \bm{p}^{\prime}\cdot\bm{k} - {\bm{p}^{\prime}}^2 \bm{p}\cdot\bm{k}}
{\left(\bm{p}^{\prime}\cdot\bm{p}\right)^{2}-\bm{p}^{2} {\bm{p}^{\prime}}^2}, \quad 
b=\frac{\bm{p}^{\prime}\cdot \bm{p}\, \bm{p} \cdot \bm{k} - \bm{p}^{2} \bm{p}^{\prime}\cdot \bm{k}}
{\left(\bm{p}^{\prime}\cdot\bm{p}\right)^{2}-\bm{p}^{2} {\bm{p}^{\prime}}^2},
    \quad 
c=\frac{\left(\bm{p}^{\prime} \times \bm{p}\right)\cdot \bm{k}}{\left|\bm{p}^{\prime} \times \bm{p}\right|^2}.	
\end{equation}

\item Contribution of planar box diagrams [the sum of diagrams $B(a)$ and $B(b)$ in Fig.~\ref{Fig:LO_TPE}]  
\begin{equation}
\begin{aligned}
 V_{B}^{(2)} &=\frac{m_N^2 g_A^4(3-2\,\bm{\tau}_1\cdot\bm{\tau}_2)}{64 f_{\pi}^4} \int\frac{d^3k}{(2\pi)^3} \Big[ X_1 + 
  X_2 \, \bm{\sigma}_1\cdot\bm{\sigma}_2
  + X_3 \, \frac{i\left(\bm{\sigma}_{1}+\bm{\sigma}_{2}\right)\cdot \bm{n}}{2}  \\
 &  + X_4 \,\left(\bm{\sigma}_{1} \cdot \bm{n}\right)\left(\bm{\sigma}_{2} \cdot \bm{n}\right) + X_5 \, \left(\bm{\sigma}_{1} \cdot \bm{q}\right)\left(\bm{\sigma}_{2} \cdot \bm{q}\right) \Big]    
\frac{1}{\varepsilon_k \varepsilon_{k+q} \omega_{k-p}^2}\left(\frac{1}{D_{B_{a}}} + \frac{1}{D_{B_{b}}}   \right),
 \end{aligned}
 \label{plD}
\end{equation}  
where the denominators corresponding to the intermediate states are given by: 	
\begin{equation}
\begin{aligned}
	D_{B_a} &= \left(E-\omega_{k-p}-\varepsilon_{k}-\omega_{p}\right) 
	\left(E-\varepsilon_{k+q}-\varepsilon_{k}-\omega_{p}-\omega_{p'}\right)
	 \left(E-\omega_{k-p}-\varepsilon_{k+q}-\omega_{p'}\right),\\ 
	D_{B_b} &= \left(E-\omega_{k-p}-\varepsilon_{k}-\omega_{p}\right)
	\left(E-\varepsilon_{k+q}-\varepsilon_{k}-\omega_{p}-\omega_{p'}\right)
	\left(E-\omega_{k-p}-\varepsilon_{k+q}-\omega_{p'}\right)\,.
\end{aligned}
\end{equation}
The coefficients $X_i$ in Eq.~(\ref{plD}) are the following functions of $a,b$ and $c$\,: 
\begin{eqnarray}
X_{1}&=&\left[\bm{k}^{2}+\bm{q}  \cdot \bm{k}\right]^{2},\quad 
X_{2}= -c^2\bm{q}^2\left[\bm{P}^{2} \bm{q}^{2}-(\bm{q} \cdot \bm{P})^{2}\right], \nonumber\\
X_{3}&=& -2(a+b)\left(\bm{k}^{2}+\left(\bm{p}^{\prime}-\bm{p}\right) \cdot \bm{k}\right),\quad 
X_{4}= -(a+b)^{2}+c^{2}\bm{q}^{2}, \nonumber\\ 
X_5 &=& c^{2}\left[\bm{P}^{2} \bm{q}^{2}-(\bm{q} \cdot \bm{P})^{2}\right],
\end{eqnarray}
with $\bm{P}=1/2(\bm{p}+\bm{p}')$.
 
\item Contribution of crossed box diagrams [the sum of diagrams $\tilde{B}(a)-\tilde{B}(f)$ in Fig.~\ref{Fig:LO_TPE}]  
\begin{equation}
\begin{aligned}
 V_{\tilde{B}}^{(2)}&=\frac{m_N^2g_A^4(3+2\bm{\tau}_1\cdot\bm{\tau}_2) }{64 f_{\pi}^4}  \int\frac{d^3k}{(2\pi)^3} \left[ X_1 - 
  X_2\, \bm{\sigma}_1\cdot\bm{\sigma}_2
 - X_4\, \left(\bm{\sigma}_{1} \cdot \bm{n}\right)\left(\bm{\sigma}_{2} \cdot \bm{n}\right) \right.\\  
 & \left. - X_5\, \left(\bm{\sigma}_{1} \cdot \bm{q}\right)\left(\bm{\sigma}_{2} \cdot \bm{q}\right) \right]\frac{1}{\varepsilon_k \varepsilon_{k+q} \omega_{p-k}\omega_{p'+k}}\\[0.5em]
& \times   \Bigg(\frac{1}{D_{\tilde{B}_{a}}} + \frac{1}{D_{\tilde{B}_{b}}} +\frac{1}{D_{\tilde{B}_{c}}} + \frac{1}{D_{\tilde{B}_{d}}}  + \frac{1}{D_{\tilde{B}_{e}}} + \frac{1}{D_{\tilde{B}_{f}}}   \Bigg),  
\end{aligned}
\end{equation}
where the denominators corresponding to intermediate states are given by:
\begin{eqnarray}
 D_{\tilde{B}_a} &= & \left(E-\omega_{k-p}-\varepsilon_{k}-\omega_{p}\right)
	\left(E-\omega_{k+p'}-\varepsilon_{k}-\omega_{p'}\right)
	\left(E-\omega_{k-p}-\omega_{k+p'}-\varepsilon_{k+q}-\varepsilon_{k}\right)	, \nonumber\\
D_{\tilde{B}_b} & = & \left(E-\omega_{k-p}-\varepsilon_{k}-\omega_{p}\right)
	\left(E-\omega_{k-p}-\varepsilon_{k+q}-\omega_{p'}\right)
	\left(E-\omega_{k-p}-\omega_{k+p'}-\varepsilon_{k+q}-\varepsilon_{k}\right) , \nonumber\\
D_{\tilde{B}_c} & = & \left(E-\omega_{k-p}-\varepsilon_{k}-\omega_{p}\right)
	\left(E-\omega_{k+p'}-\varepsilon_{k}-\omega_{p'}\right)
	\left(E-\varepsilon_{k+q}-\varepsilon_{k}-\omega_{p}-\omega_{p'}\right) , \\
D_{\tilde{B}_d} &= & \left(E-\omega_{k+p'}-\varepsilon_{k+q}-\omega_{p}\right)
\left(E-\omega_{k-p}-\varepsilon_{k+q}-\omega_{p'}\right)
		\left(E-\omega_{k-p}-\omega_{k+p'}-\varepsilon_{k+q}-\varepsilon_{k}\right) , \nonumber\\ 
D_{\tilde{B}_e} &= & \left(E-\omega_{k+p'}-\varepsilon_{k+q}-\omega_{p}\right) \left(E-\omega_{k+p'}-\varepsilon_{k}-\omega_{p'}\right)
	\left(E-\omega_{k-p}-\omega_{k+p'}-\varepsilon_{k+q}-\varepsilon_{k}\right) , \nonumber\\
D_{\tilde{B}_f} & = & \left(E-\omega_{k+p'}-\varepsilon_{k+q}-\omega_{p}\right) \left(E-\omega_{k-p}-\varepsilon_{k+q}-\omega_{p'}\right)
	\left(E-\varepsilon_{k+q}-\varepsilon_{k}-\omega_{p}-\omega_{p'}\right).
\nonumber
\end{eqnarray}
\end{itemize}

 \begin{figure}[t]
  \includegraphics[width=0.6\textwidth]{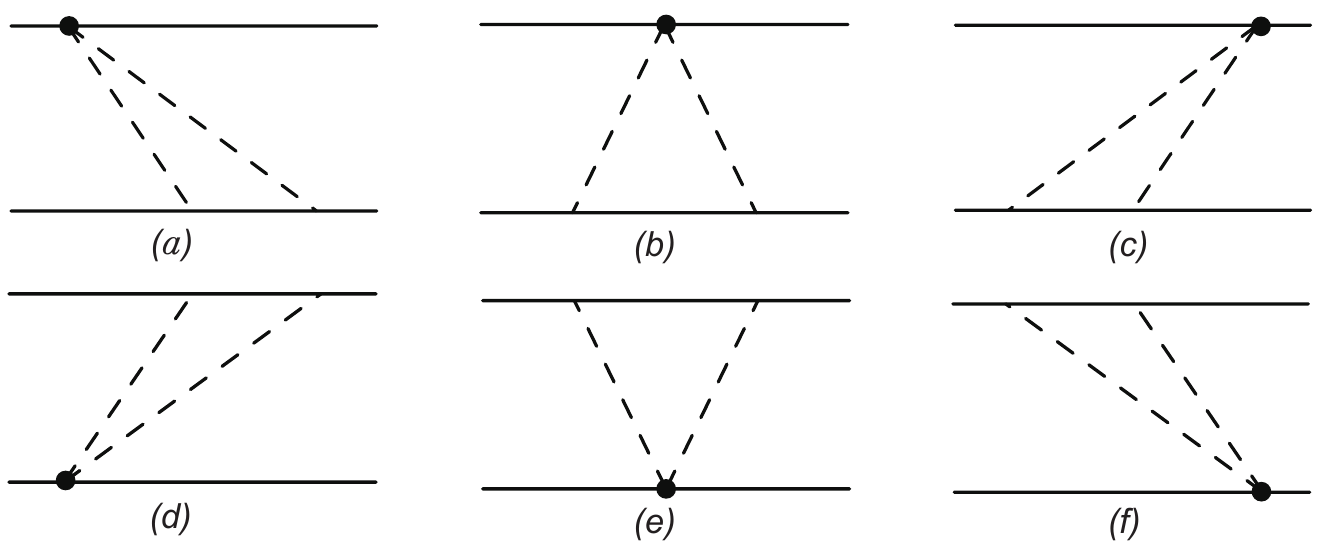}
  \caption{Time-ordered diagrams contributing to the TPE potential at NNLO. Solid and dashed lines correspond to nucleons and pions, respectively. The filled circles denote the vertices from $\mathcal{L}_{\pi N}^{(2)}$.}
  \label{Fig:NLO_TPE}
\end{figure}

The diagrams giving non-vanishing contributions to the TPE potential at third order ($\nu=3$, NNLO) in our calculations are shown in Fig.~\ref{Fig:NLO_TPE}, 
 where the football diagrams do not contribute since we keep only the LO terms in the expansion of the Dirac spinors. Then, using the TOPT rules we obtain the following contribution 
\begin{equation}
\begin{aligned} 
  V_{2\pi}^{(\nu=3)} & = \frac{3m_N\,g_A^2}{16 f_\pi^4} \int \frac{d^3 k}{(2\pi)^3} \left[(\bm{k}^2+(\bm{p}'-\bm{p})\cdot\bm{k}) - (a+b)\frac{i(\bm{\sigma}_1+\bm{\sigma}_2)\cdot\bm{n}}{2} \right] \frac{1}{\varepsilon_k\varepsilon_{k+q}\omega_{p-k}}\\
  & \times \Bigg\{ \left[ 4 c_1 M_\pi^2 -\frac{c_2}{m_N^2} \bigg( \bm{p}\cdot\bm{k}\,\bm{p}\cdot(\bm{k}+\bm{q}) + \bm{p}'\cdot\bm{k}\,\bm{p}'\cdot(\bm{k}+\bm{q}) \bigg) + 2c_3 \bm{k}\cdot(\bm{k}+\bm{q}) \right] \\
  & \times \Bigg(\frac{1}{D_{a}} + \frac{1}{D_{b}} + \frac{1}{D_{c}} + \frac{1}{D_{d}} + \frac{1}{D_{e}} + \frac{1}{D_{f}}\Bigg) \\
  &  + \left[ \frac{c_2}{m_N^2}\varepsilon_k\varepsilon_{k+q}(\omega_p+\omega_{p'}) + 2 c_3 \varepsilon_k\varepsilon_{k+q}\right] 
   \Bigg(\frac{1}{D_{a}} - \frac{1}{D_{b}} + \frac{1}{D_{c}} + \frac{1}{D_{d}} - \frac{1}{D_{e}} + \frac{1}{D_{f}}\Bigg)\\
  & + \left[ \frac{c_2}{m_N^2} \varepsilon_k  \left( \omega_p \bm{p}\cdot(\bm{k}+\bm{q}) + \omega_{p'}\bm{p}'\cdot(\bm{k}+\bm{q}) \right) \right] 
   \Bigg(\frac{1}{D_{a}} - \frac{1}{D_{b}} - \frac{1}{D_{c}} + \frac{1}{D_{d}} - \frac{1}{D_{e}} - \frac{1}{D_{f}}\Bigg)\\
  &  - \left[ \frac{c_2}{m_N^2} \varepsilon_{k+q}  \left( \omega_p \bm{p}\cdot\bm{k} + \omega_{p'}\bm{p}'\cdot\bm{k} \right) \right]    
   \Bigg(\frac{1}{D_{a}} + \frac{1}{D_{b}} - \frac{1}{D_{c}} + \frac{1}{D_{d}} + \frac{1}{D_{e}} - \frac{1}{D_{f}}\Bigg) \Bigg\} \\[1em]
&+\frac{c_4m_N g_A^2\bm{\tau}_1\cdot\bm{\tau}_2}{8 f_{\pi}^4} \int\frac{d^3k}{(2\pi)^3} \Big[
  X_2 \bm{\sigma}_1\cdot\bm{\sigma}_2
  + \frac{X_3}{2} \frac{i\left(\bm{\sigma}_{1}+\bm{\sigma}_{2}\right)\cdot \bm{n}}{2}  
   + X_4 \left(\bm{\sigma}_{1} \cdot \bm{n}\right)\left(\bm{\sigma}_{2} \cdot \bm{n}\right) \\
& + X_5 \left(\bm{\sigma}_{1} \cdot \bm{q}\right)\left(\bm{\sigma}_{2} \cdot \bm{q}\right) \Big]   
 \frac{1}{ \varepsilon_{k+q} \omega_{p-k}} \Bigg(\frac{1}{D_a} + \frac{1}{D_b} + \frac{1}{D_c} + \frac{1}{D_d} + \frac{1}{D_e} + \frac{1}{D_f}    \Bigg),
\end{aligned}
\end{equation} 
where the denominators corresponding to the intermediate states in diagrams of  Figs.~\ref{Fig:NLO_TPE} (a)-(f) are the same as the ones given in Eq.~(15), i.e.
\begin{equation}
	D_a  = D_{T_a}, \quad D_b = D_{T_b}, \quad D_c = D_{T_c}, \quad
	D_d  = D_{\tilde{T}_a}, \quad D_e = D_{\tilde{T}_b}, \quad D_f = D_{\tilde{T}_c}.
\end{equation}

To renormalize the one-loop diagrams we apply subtractive renormalization by expanding the integrands in powers of the external momenta and the pion mass and subtracting those contributions which lead to 
divergent and power counting violating contributions in the removed cutoff limit. We checked explicitly that all these subtraction terms are indeed cancelled by appropriate local counter terms of the 
NN contact interaction Lagrangian. The renormalized contributions of
the TOPT diagrams are calculated numerically.

\subsubsection{Once-iterated one-pion exchange} 

Using the previous diagrammatic rules of TOPT, we obtain the OPE potential	
\begin{equation}
	V_{\mathrm{OPE}} =-\frac{g_{A}^{2}}{4 f_{\pi}^{2}} \, \bm{\tau}_{1} \cdot \bm{\tau}_{2} \,  \frac{1}{\varepsilon_q}\frac{\left(\bar{u}_{3} \gamma^{\mu} \gamma_{5} q_{\mu} u_{1}\right)\left(\bar{u}_{4} \gamma^{\nu} 
	\gamma_{5} q_{\nu} u_{2}\right)}{\omega_p+\omega_{p'}+\varepsilon_q-E-i\epsilon}\,.
\end{equation}
Apparently, when dealing with the once-iterated OPE potential	
\begin{equation}
  V_\mathrm{OPE}\,  G\, V_\mathrm{OPE}  = \int\frac{d^3 k}{(2\pi)^3}\frac{m_N^2}{\bm{k}^2+m_N^2} \frac{V_\mathrm{OPE}(p',k)\,V_\mathrm{OPE}(k,p)}{E-2\,\omega_k +i\epsilon}\,, 	
\end{equation}
where the energy denominator is just  the two-nucleon Green function of the Kadyshevsky equation,\footnote{
Within our TOPT approach, one can cast the Kadyshevsky equation into other forms of the scattering equation, however this requires to change the potential accordingly.}, 
which is obtained directly from our TOPT rules, one encounters the poles in the denominators of the half-off-shell OPE potentials $V_\mathrm{OPE}(p',k)$ and $V_\mathrm{OPE}(k,p)$.
To avoid this technical complication it is convenient to eliminate the energy-dependence of the OPE potential $V_\mathrm{OPE}(p,p')$ by performing an expansion in powers of $E-\omega_p-\omega_{p'}$,  
and obtain, up to the accuracy of our calculation, an equivalent
energy-independent potential $V_\mathrm{\slashed E}  =
V_{\mathrm{OPE},\slashed E}^{(0)} + V_{2\pi,\slashed E}^{(2)}$,  which satisfies 
\begin{equation}
V_\mathrm{OPE} + V_\mathrm{OPE}  \,G\, V_\mathrm{OPE} =  V_{\mathrm{OPE},\slashed E}^{(0)} +V_{2\pi, \slashed E}^{(2)} + V_{\mathrm{OPE}, \slashed E}^{(0)}  \,G\, V_{\mathrm{OPE}, \slashed E} + \mathcal{O}(\nu=4),
\end{equation}
where the energy-independent OPE potential is given by
\begin{equation}\label{Eq:Eint_OPE}
V_{\mathrm{OPE},\slashed E}^{(0)} 
= -\frac{g_{A}^{2}\bm{\tau}_{1} \cdot \bm{\tau}_{2}}{4 f_{\pi}^{2}}  \frac{1}{\varepsilon_q^2} 
 \left(\bar{u}_{3} \gamma_{\mu} \gamma_{5} q^\mu u_{1}\right) \left(\bar{u}_{4} \gamma_{\nu} \gamma_{5} q^{\nu} u_{2}\right), 
\end{equation}
and, for simplicity, we keep the full form of Dirac spinors of the nucleon, which is equivalent to including also higher order contributions of the OPE potential.
The second term $V_{2\pi,\slashed E}^{(2)}$ is written as
\begin{equation}
\begin{aligned}
V_{2\pi, \slashed E}^{(2)} 
=& \frac{1}{2}  \left(\frac{g_{A}^{2}}{4 f_{\pi}^{2}} \right)^2 (3- \bm{\tau}_{1} \cdot \bm{\tau}_{2})
 \int \frac{d^3k}{(2\pi)^3} \frac{m_N^2}{\bm{ k}^2+m_N^2}  \frac{\varepsilon_{p'-k} +\varepsilon_{p-k}}{\varepsilon_{p'-k}^3 \,\varepsilon_{p-k}^3} \\
& \times \left[ \bm{\sigma}_1\cdot(\bm{p}'-\bm{k})\bm{\sigma}_1\cdot(\bm{k}-\bm{p})\right] \, \left[\bm{\sigma}_2\cdot(\bm{p}'-\bm{k})\bm{\sigma}_2\cdot(\bm{k}-\bm{p})\right],
\end{aligned}
\end{equation}
where we keep only the LO terms in the expansion of the Dirac spinors, similarly to the treatment of TPE potential in the last subsection. 
Notice that the $V_{2\pi, \slashed E}^{(2)}$ is obtained by cancelling the
Kadyshevsky denominator and is the part of the TPE potential at NLO, which has the same spin structure as the
planar box diagram of the same order.   

According to the above discussion, the once-iterated OPE potential leads to   
\begin{equation}
 V_{2\pi, it}(p',p) = V_{\mathrm{OPE},\slashed E}^{(0)}  \,G\, V_{\mathrm{OPE},\slashed E}^{(0)} = \int\frac{d^3k}{(2\pi)^3}\frac{m_N^2}{\bm{k}^2+m_N^2} \frac{V^{(0)}_{\mathrm{OPE},\slashed E}(p',k)\,V^{(0)}_{\mathrm{OPE},\slashed E}(k,p)}{E-2\,\omega_k+i\epsilon}\,. 	
\end{equation}
We found that $V^{(0)}_\mathrm{OPE}(p',p)$ has a milder ultraviolet behavior compared to its non-relativistic analogue. In particular for fixed $p'$ and large $p$ we have $V_\mathrm{OPE}^{(0)} \sim 1/p$. The benefit comes to the once-iterated OPE potential, 
which is UV convergent for all partial waves. This allows us to directly calculate the integral by using the standard Gauss-Legendre quadratures. While the non-relativistic counterpart of the once-iterated OPE potential is linearly divergent, 
one can use the dimensional regularization to obtain a closed finite form, as done in Ref.~\cite{Kaiser:1997mw}. 

\subsubsection{NN $T$-matrix  and phase shifts}
Finally, we obtain the $T$-matrix of NN scattering at one-loop order in the Born expansion 
\begin{equation}
\begin{aligned}
	T(p',p) &= V_{\mathrm{OPE},\slashed E}^{(0)}(p',p) + V_{2\pi, \slashed E}^{(2)}(p',p) + V_{2\pi, irr}^{(2)}(p',p) + V_{2\pi, irr}^{(3)}(p',p) + V_{2\pi, it}(p',p) \nonumber\\
	&= V_{\mathrm{OPE},\slashed E}^{(0)}(p',p) + V_{2\pi, irr}(p',p) + V_{2\pi, it}(p',p),
\end{aligned}
\end{equation}
where $V_{2\pi, irr}$ refers to the NN potential as the sum of all two-nucleon irreducible TPE contributions up to NNLO: $V_{2\pi, irr}=V_{2\pi, \slashed E}^{(2)} + V_{2\pi, irr}^{(2)}+V_{2\pi, irr}^{(3)}$. 

We follow the steps given in Ref.~\cite{Erkelenz:1971caz} and perform the partial wave decomposition of the NN $T$-matrix to express it in the standard $lsj$ representation. 
Then, the phase shifts and mixing angles can be perturbatively calculated via~\cite{Gasser:1990ku,Kaiser:1997mw} 
\begin{equation}
\begin{aligned}
\delta_{l}^{sj} &=-\frac{p\, m_N^2}{16 \pi^2 \sqrt{p^2+m_N^2}} \operatorname{Re}\langle lsj|T|lsj\rangle, \\
\epsilon_{j} &=\frac{p\, m_{N}^{2}}{16 \pi^{2} \sqrt{p^{2}+m_{N}^{2}}} \operatorname{Re}\langle j-1,1, j|T| j+1,1, j\rangle,
\end{aligned}
\label{unitarized}
\end{equation}
where $p \equiv |\bm{p}|$.

 Notice that a unique one-to-one correspondence between the scattering
 amplitude and the corresponding phase shifts only exists in case if the calculated amplitude is exactly unitary.
 If a perturbatively calculated amplitude is not exactly unitary, then the corresponding phase shifts depend on the method of unitarizing. 
 Non-unitary amplitudes can be unitarized by changing their real parts, imaginary parts or both.
 There is an (continuously) infinite number of ways such unitarizations
 can be accomplished. All of them are equally good provided the
 changes made to the amplitude are of a higher order. 
 The phase shifts corresponding to different unitarization approaches differ by contributions of higher orders. In our case, the perturbative amplitude is not exactly unitary and, therefore, we apply a technically convenient 
 method of unitarization specified in Eq.~(\ref{unitarized}).
  When expanding the calculated phase
  shifts in powers of the EFT expansion parameter, 
$\delta = \delta^{(\nu=0)}+\delta^{(\nu=2)}+\delta^{(\nu=3)}+
 \delta^{\mathcal{O}(\nu=3)}$, only the terms up to NNLO, i.e.~$\delta^{(\nu=0,2,3)}$, are uniquely determined and
 independent of the unitarization procedure, while the induced higher-order contributions
 $\delta^{\mathcal{O}(\nu=3)}$ are ambiguous. Given the smallness
 of the calculated phase shifts in  $F$- and higher
partial waves, the dependence of our results on the
unitarization procedure is negligible in those channels.   
 
\section{Peripheral phase shifts} 
\label{Section3} 
In this section we calculate the NN phase shifts and mixing angles for
partial waves with the angular momenta $l\geq 2$ and $j\geq 2$, and
compare them with the corresponding results of the non-relativistic
approach. We start by first performing the consistency checks of our
results for the TPE potential $V_{2\pi, irr}$ up to NNLO and the
once-iterated OPE, $V_{2\pi, it}$, by taking the large $m_N$ limit.  

The values of parameters used in the following calculations are as follows: 
the average pion and nucleon masses $M_\pi=138$ MeV and $m_N=938.918$
MeV, the pion decay constant $f_\pi=92.4$ MeV is fixed to its physical value; 
the axial coupling $g_A$ fixed as $1.267$ for LO calculation, and
changed to $1.29$ to account for the Goldberger-Treiman discrepancy at NLO and NNLO. 
For our NNLO calculations, we also need to specify 
the numerical values of the LECs $c_1$, $c_2$, $c_3$, and $c_4$. 
We take $c_1=-0.74$ GeV$^{-1}$, $c_2=1.81$ GeV$^{-1}$, $c_3=-3.61$
GeV$^{-1}$, and $c_4=2.17$ GeV$^{-1}$ obtained from the order-$Q^2$
matching to the $\pi N$ subthreshold parameters, determined by the
Roy-Steiner analysis of $\pi N$ scattering~\cite{Hoferichter:2015hva},
using the covariant formulation of ChEFT~\cite{Siemens:2016jwj}. We use the same values for parameters to obtain the NNLO results of the non-relativistic ChEFT.  

 \begin{figure}[b]
\includegraphics[width=0.8\textwidth]{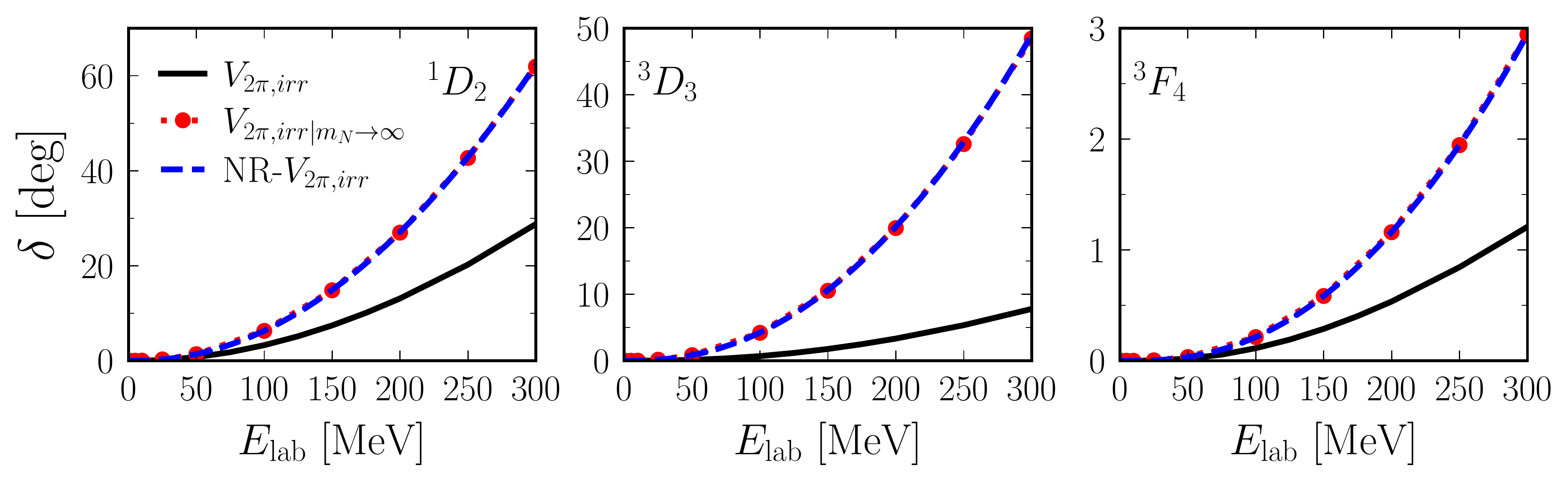}
  \caption{The phase shifts of $^1D_2$, $^3D_3$ and $^3F_4$ partial waves from the TPE potential 
  up to NNLO. 
  The black solid lines denote our results with the physical nucleon mass, the red dotted lines are generated by taking $m_N$ in TPE potential 
  to infinity (numerically we take $m_N=1000\, m_N^\mathrm{Phys}$). 
  The blue dashed lines are the results  of non-relativistic TPE potential.}
  \label{Fig:TPE_check}
\end{figure}  

\subsection{Consistency checks of the two-pion exchange contributions}

For $D$, $F$ and higher partial waves, there are no contact-interaction contributions to the potential at NLO and NNLO. 
Thus, our irreducible TPE potential obtained using the subtractive
renormalization and the non-relativistic TPE potential calculated using the dimensional 
regularization should give the same results when the nucleon mass $m_N$ is taken to infinity. 
This can be verified numerically with good accuracy by calculating the
phase shifts in both approaches for large values of the nucleon mass.

In Fig.~\ref{Fig:TPE_check} we present some typical phase shifts of $D$ and $F$ waves given by the irreducible TPE diagrams in our scheme for the physical nucleon mass $m_N^\mathrm{Phys}$ and for a large nucleon mass $m_N=1000\,m_N^\mathrm{Phys}$. 
Our results are consistent with the ones of the non-relativistic TPE
potential. Notice that our TPE contributions up to NNLO for the
physical value of the nucleon mass are smaller in magnitude than their non-relativistic analogues, particularly for  the $^3D_3$ partial wave.

\begin{figure}[t]
  \includegraphics[width=0.8
  \textwidth]{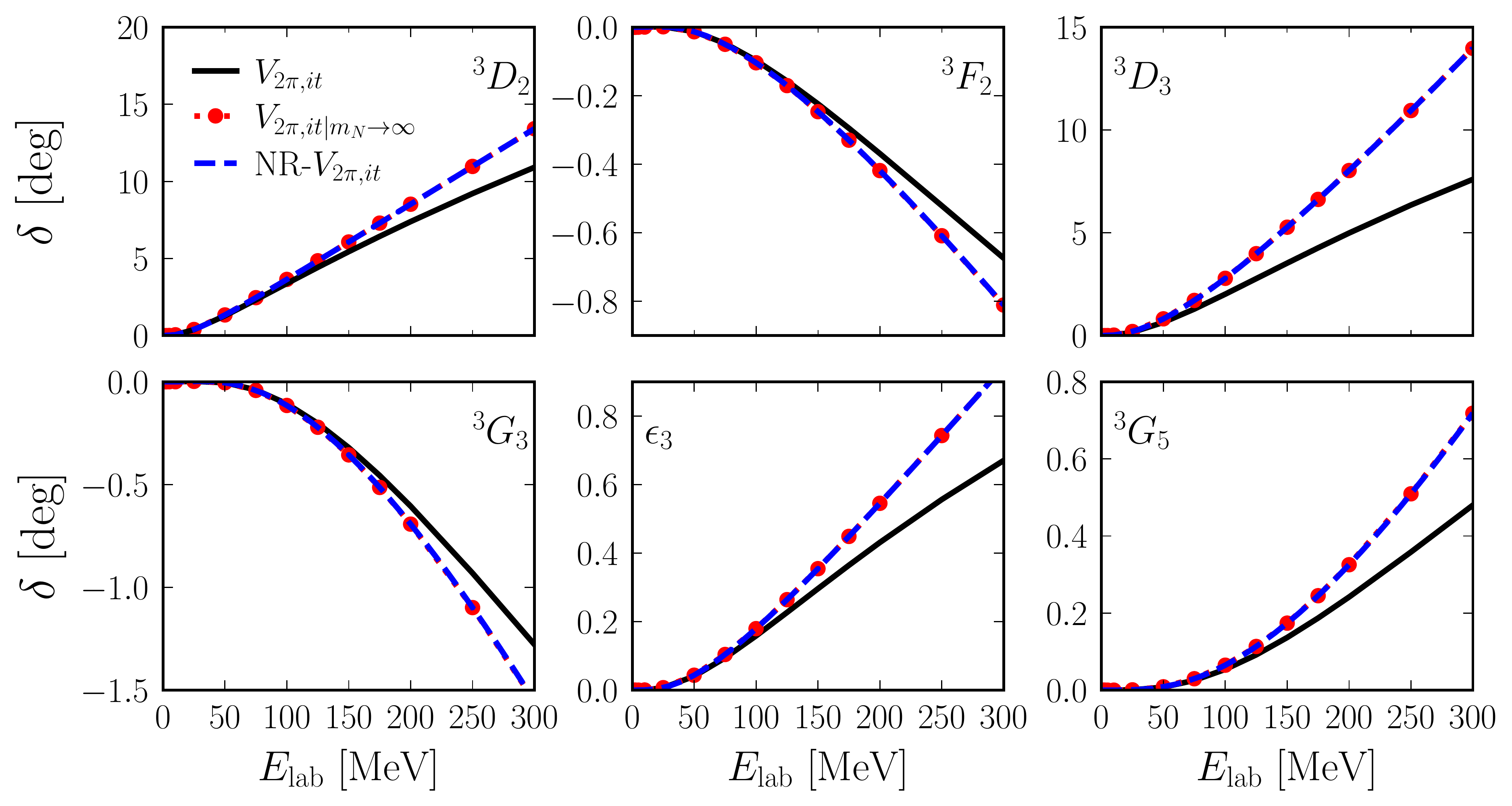}
  \caption{Partial wave phase shifts corresponding to the once-iterated OPE potential. The black solid lines denote our results with the physical nucleon mass, the red dotted lines are generated by taking $m_N$ in $V_{2\pi, it}$ to infinity (numerically we take $m_N=1000\, m_N^\mathrm{Phys}$). The blue dashed lines are the results of non-relativistic once-iterated OPE potential.}
  \label{Fig:VGV_check}
\end{figure}

Furthermore, we also present the consistency check of the once-iterated OPE contribution in Fig.~\ref{Fig:VGV_check}. where the phase shifts of several partial waves receiving sizable contributions from $V_{2\pi, it}$ are shown. 
 To check the reliablility of our numerical evaluation, we take $m_N$ in $V_{2\pi, it}$ 
 to infinity and indeed reproduce the non-relativistic results obtained using 
 the analytic expressions of the once-iterated OPE potential, as specified by Eqs.~(31)-(34) in Ref.~\cite{Kaiser:1997mw}. 
 Notice that the relativistic correction factor $m_N/\sqrt{m_N^2+p^2}$ included in Eq.~(24) of  Ref.~\cite{Kaiser:1997mw}   
 needs to be removed for the purposes of our comparison. We can see that for the physical nucleon mass the phase shifts obtained using our $V_{2\pi, it}$ are smaller than the ones of the non-relativistic case, 
 particularly for the $^3D_3$ partial wave. That is because the once-iterated OPE potential is slightly less attractive/repulsive than its non-relativistic counterpart.

\subsection{Peripheral phase shifts}

Below we present our results for the partial wave phase shifts for $l \geq 2$ and $j\geq 2$ using our chiral potential up to NNLO. 

\subsubsection{D-waves}
\begin{figure}[t]
  \includegraphics[width=0.8\textwidth]{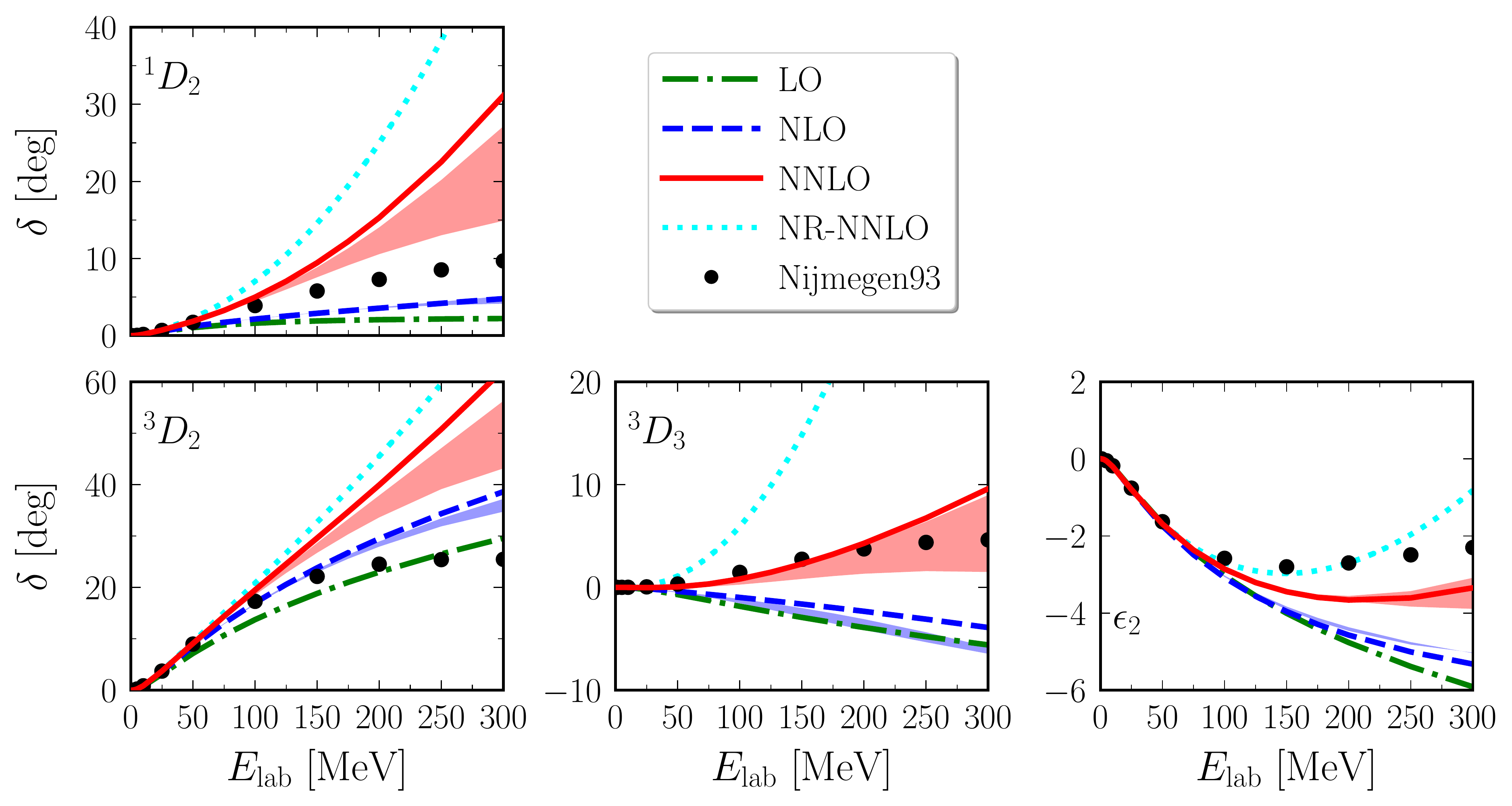}
  \caption{$D$-wave neutron-proton phase shifts and the mixing angle $\epsilon_2$ for laboratory 
  energies below $300$ MeV. The green dot-dashed curve is the LO
  result (which corresponds to the pure OPE), the blue dashed and red solid lines denote the NLO and NNLO results, respectively. 
  The non-relativistic results at NNLO are shown as the cyan dotted
  lines. The filled circles represent the results of the Nijmegen
  partial wave analysis \cite{Stoks:1993tb}. 
  The light-blue and red bands correspond to the NLO and NNLO
  potentials with the loop integrals in the TPE diagrams regulated using the cutoff $\Lambda=500-800$ MeV.}
  \label{Fig:Dwave}
\end{figure}

Our results for the D-wave phase shifts and the mixing angle $\epsilon_2$ are presented in Fig.~\ref{Fig:Dwave}. 
The LO result (corresponding to the pure OPE potential), shown by the green dot-dashed curves, provides the major contributions to $^3D_2$ and $\epsilon_2$ but is too weak in the $^1D_2$ channel 
and gives the opposite trend in comparison with the empirical phase shifts for the $^3D_3$ partial wave. Including the NLO correction obtained using the subtractive renormalization gives the 
correct direction of improvement for all $D$-wave phase shifts, as shown by the blue dashed lines. 
However its contribution is relatively small, which is (partly) due to the strong cancelation between the leading TPE potential and the once-iterated OPE. For example, 
the contributions of $V_{2\pi, it}$ to the $^3D_2$ and $^3D_3$ partial
waves are quite large, as seen in Fig.~\ref{Fig:VGV_check}, but adding
them to the contributions of the strongly repulsive TPE potential  
results in small attractive contributions.   

At NNLO (results shown by the red solid lines), using the subtractive
renormalization we found sizable improvement for the $^3D_3$ phase shifts and $\epsilon_2$. On the other hand, the good agreement with the data observed at LO and NLO 
is notably worsened for the $^1D_2$ and $^3D_2$ partial waves for energies $E_\mathrm{lab}>100$ MeV. In comparison with the non-relativistic 
NNLO results (shown by the dotted lines), which are exploding beyond $E_\mathrm{lab}>50$ MeV for all $D$ waves, 
the improvement delivered by our approach is visible, particularly for the $^3D_3$ partial wave. As noticed in Ref.~\cite{Epelbaum:2003gr}, 
 strong disagreement with the empirical data is caused by unphysical short-distance components of the non-relativistic TPE potential. 
Following the suggestion of that work to use an alternative
regularization scheme instead of the dimensional regulartization, we
apply the cutoff regularization with the cutoff $\Lambda$ varying from
$500$ MeV to $800$ MeV to loop integrals in TPE diagrams. 
The corresponding NLO and NNLO results for phase shifts are presented as the light-blue and red bands in Fig.~\ref{Fig:Dwave}. 
For the total angular momentum $j=2$, i.e. in $^1D_2$ and $^3D_2$
partial waves, the improvement is visible. For the $^3D_3$ phase shift
and the $\epsilon_2$ mixing angle the results are similar to the
previous ones using the subtractive renormalization in the limit of a  removed regulator. 
This observation is in line with our expectations since the calculated
TPE potential has a milder ultraviolet behavior and the short-distance contribution of loop integrals is suppressed in comparison with the non-relativistic case. 
Note that we do not show the $^3D_1$ phase shift due to the strong coupling between the $^3D_1$ and $^3S_1$ channels.    

\begin{figure}[b]
  \includegraphics[width=0.8\textwidth]{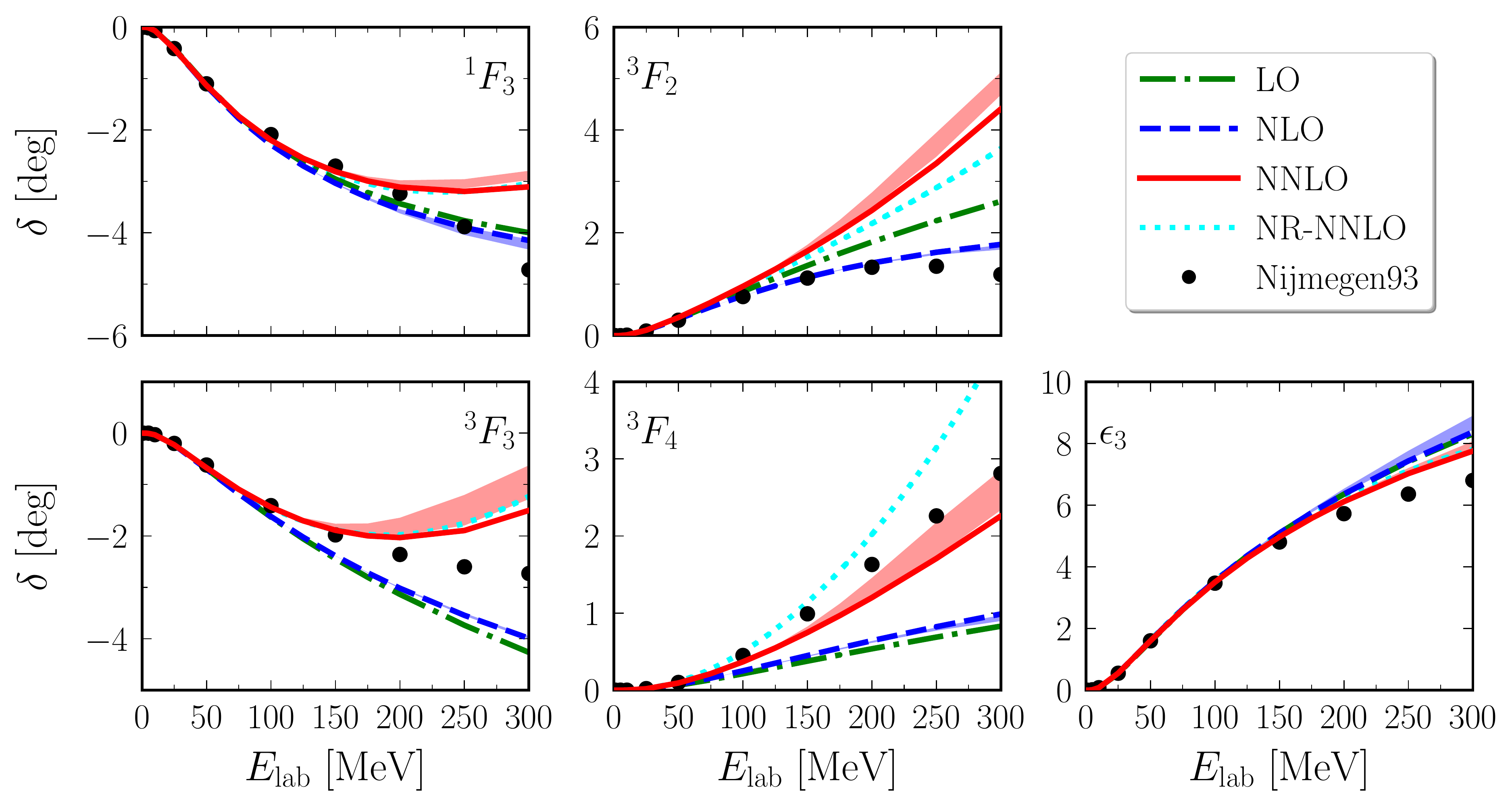}
  \caption{$F$-wave neutron-proton phase shifts and the mixing angle $\epsilon_3$ for laboratory 
  energies below $300$ MeV. Notations are as in Fig.~\ref{Fig:Dwave}.}
  \label{Fig:Fwave}
\end{figure}

\subsubsection{F-waves}
 Differently from the case of the $D$-waves, the empirical phase shifts for $F$-waves are quite small (less than $5^\circ$).  Our results for the $F$-wave phase shifts and mixing angle $\epsilon_3$ are shown in Fig.~\ref{Fig:Fwave}.
The LO results for $^1F_3$, $^3F_2$, $^3F_3$ and $\epsilon_3$ roughly
match the data, while the OPE contribution for $^3F_4$ is very
small. The inclusion of the NLO terms leads mainly to small corrections, 
however visible improvement is seen in the $^3F_2$ partial wave. 
Up to NNLO our results for the $^1F_3$, $^3F_3$, and $^3F_4$ partial
waves are in a good agreement with the data up to $E_\mathrm{lab}=150$ MeV.  For larger energies, the NNLO correction becomes too strong due to the subleading TPE potential.
Similar behavior is also observed for the non-relativistic NNLO results. In the $^3F_2$ partial wave, the correct tendency achieved at NLO is altered by including the subleading TPE potential. 
Furthermore, we also present the NLO and NNLO results with the cutoff regularization of the loop integrals in the TPE potential. They are similar to the results of using the subtractive renormalization with removed regulator limit. 
This indicates that the short-distance components in the TPE are rather small in our NNLO results for $F$-waves.       
Last but not least, we emphasize that the remaining discrepancies between the calculated
and empirical $F$-wave phase shifts are comparable with the natural-size 
contributions of the leading contact interactions, which appear at
sixth order in the EFT expansion \cite{Reinert:2017usi,Epelbaum:2019kcf}.  

\subsubsection{G-waves}
\begin{figure}[t]
  \includegraphics[width=0.8\textwidth]{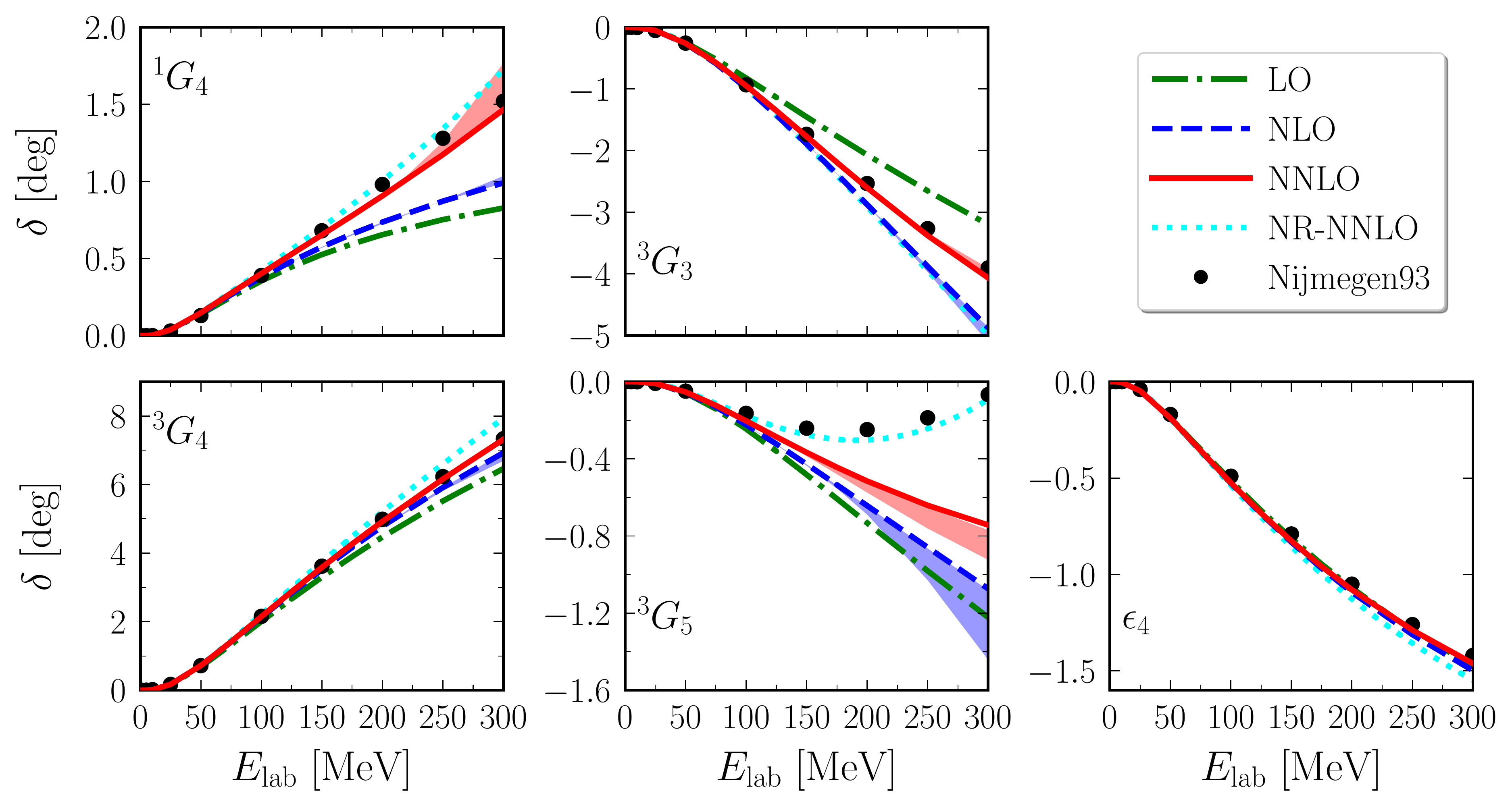}
  \caption{$G$-wave neutron-proton phase shifts and the mixing angle $\epsilon_4$ for laboratory 
  energies below $300$ MeV. Notations are as in Fig.~\ref{Fig:Dwave}.}
  \label{Fig:Gwave}
\end{figure}

The $G$-wave phase shifts and the mixing angle $\epsilon_4$ up to NNLO
are shown in Fig.~\ref{Fig:Gwave}. Our NNLO results describe the
Nijmegen partial wave analysis  rather well, except for the $^3G_5$ partial wave. 
The convergence pattern of the chiral expansion for the $^3G_3$,
$^3G_4$ partial waves and for $\epsilon_4$ looks very reasonable with
the LO potential giving the dominant contribution and the corrections due to the 
TPE potential being quite small.  As for the $^1G_4$ phase shifts, the
NNLO correction of the TPE potential provides a sizable contribution
leading to a good agreement with data. In comparison with the NNLO results of non-relativistic ChEFT, 
our approach gives a slightly better description for $^1G_4$, $^3G_3$, $^3G_4$ partial waves and for $\epsilon_4$. 

The situation is different for the $^3G_5$ partial wave, where the
non-relativistic result shows a rather good agreement with the data at
NNLO. This agreement is, however, accidental. In particular, the 
relativistic corrections to the TPE potential $\propto c_i/m_N$, which in the
non-relativistic counting scheme appear at fifth chiral order (i.e., at
N$^4$LO), are of the same size as the difference between the NNLO and
NR-NNLO lines in Fig.~\ref{Fig:Gwave}, see Ref.~\cite{Entem:2014msa}.
These contributions are already
taken into account in our NNLO results along with an infinite set
of $c_i/m_N^n$,  $n\ge 2$, corrections. Thus, we expect the
convergence of the covariant chiral EFT approach for this
partial wave to be superior as compared to the non-relativistic
framework.  It is also worth emphasizing that the empirical
phase shifts in the $^3G_5$ channel reflect a subtle interplay between
a repulsive long-range and attractive short-range interactions
and appear to be much smaller than those in other $G$-waves. It is,
therefore, not straightforward to draw conclusions about the
convergence of the chiral expansion in this particular partial wave. 
We also note in this context that the small NLO correction we found 
in this channel is due to a cancelation between the individually
larger contributions of the leading TPE potential and the
once-iterated OPE potential.

For $G$-waves, the results with the cutoff regularization of the loop integrals in the TPE potential are similar to those using the subtractive renormalization with removed regulator limit.           

\subsubsection{H-waves}
\begin{figure}[t]
  \includegraphics[width=0.8\textwidth]{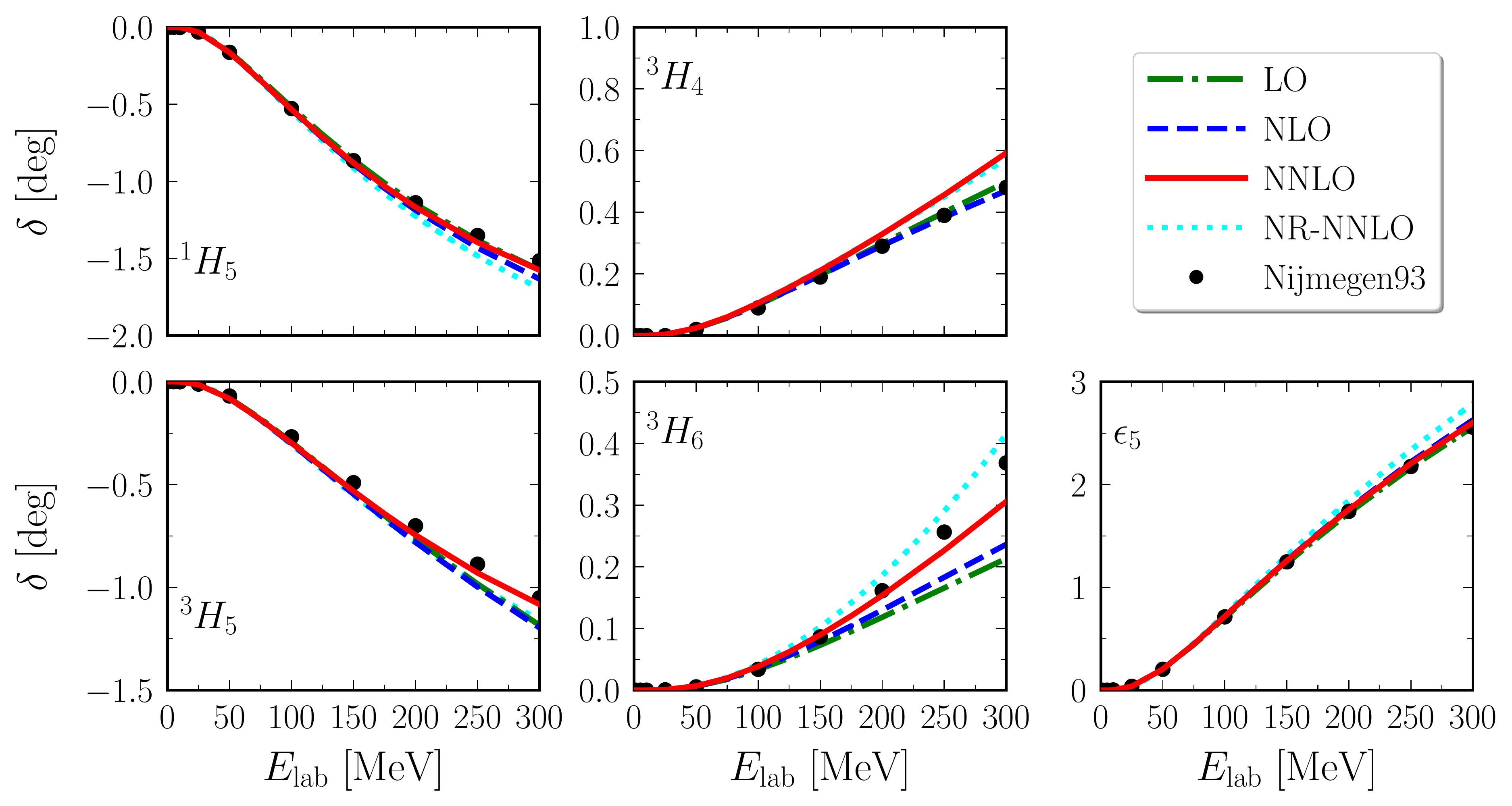}
  \caption{$H$-wave neutron-proton phase shifts and the mixing angle $\epsilon_5$ for laboratory 
  energies below $300$ MeV. Notations are as in Fig.~\ref{Fig:Dwave}.}
  \label{Fig:Hwave}
\end{figure}

In Fig.~\ref{Fig:Hwave} we present the $H$-wave phase shifts and the
mixing angle $\epsilon_5$. Basically, the OPE potential can achieve a
rather good description of Nijmegen data, except for the $^3H_6$
partial wave. The corrections due to the TPE  
are quite small. 
For the $^3H_5$ and $^3H_6$ phase shifts, the subleading TPE potential
is needed to describe the empirical phase shifts for
$E_\mathrm{lab}>150$ MeV. The NNLO results of $^3H_6$ phase shifts are
slightly lower than the empirical data, while the NNLO potential gives
very small positive contributions for the $^3H_4$ phase
shifts. However such differences are insignificant due to the small
empirical values of the $^3H_4$ and $^3H_6$ phase shifts (less than
$0.5^\circ$). In comparison with the NNLO results of the
non-relativistic scheme, a different tendency is   seen in the $^3H_6$
partial wave. The non-relativistic NNLO result overshoots the
empirical phase shifts, while our NNLO result lies below the data
points. For the other $H$-waves and the mixing angle $\epsilon_5$, our approach gives a slightly better description than the non-relativistic formalism.

\subsubsection{I-waves}
\begin{figure}[t]
  \includegraphics[width=0.8\textwidth]{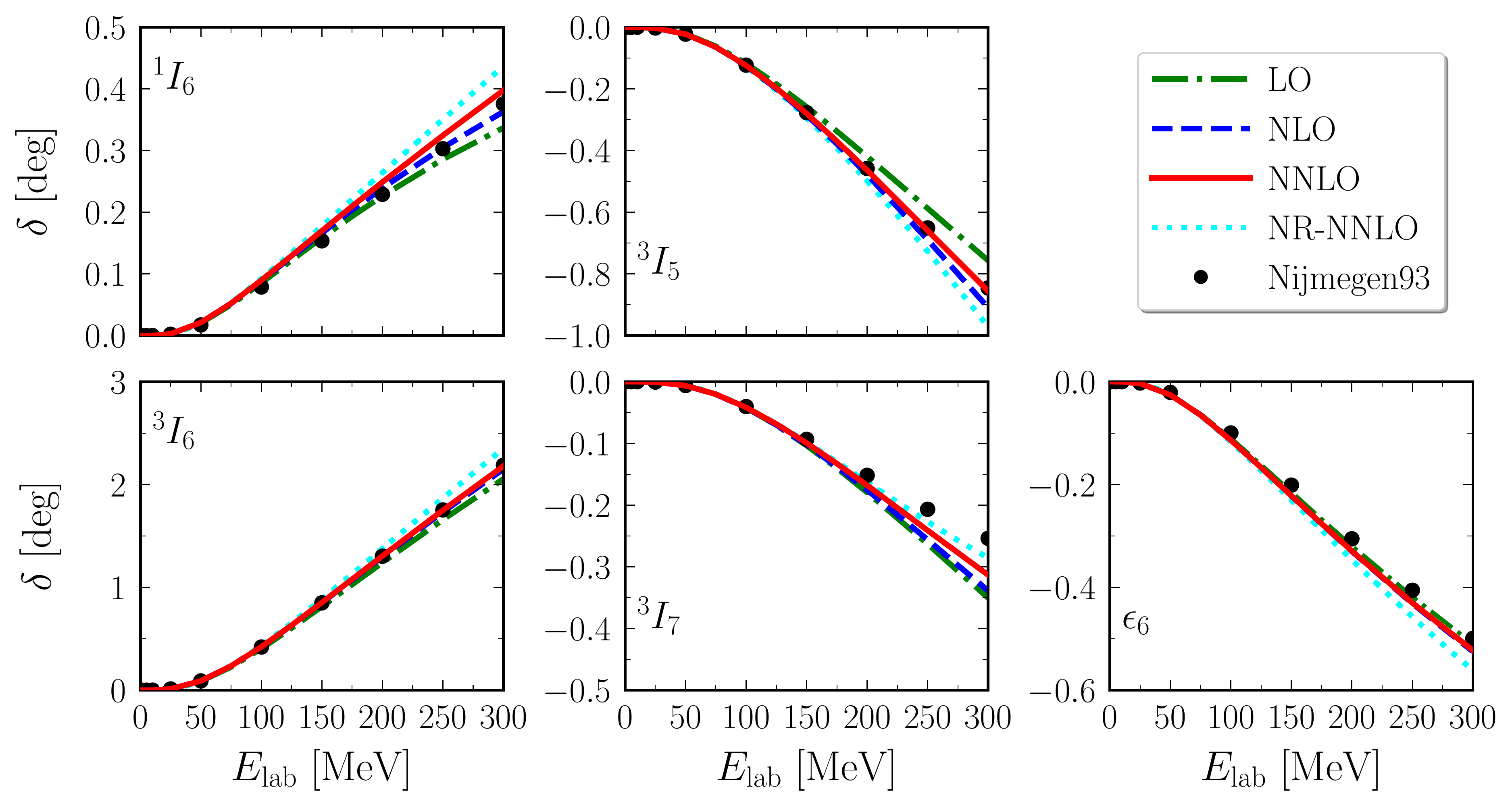}
  \caption{$I$-wave neutron-proton phase shifts and the mixing angle $\epsilon_6$ for laboratory 
  energies below $300$ MeV. Notations are as in Fig.~\ref{Fig:Dwave}.}
  \label{Fig:Iwave}
\end{figure}

The $I$-wave phase shifts and mixing angle $\epsilon_6$ are shown in
Fig.~\ref{Fig:Iwave}, where the pure OPE potential already provides a
very good approximation. 
Although the NLO and NNLO corrections  are relatively small, their
contributions are visible and slightly improve the description  
of Nijmegen data with $E_\mathrm{lab}>200$ MeV, particularly for the
$^1I_6$, $^3I_5$, and $^3I_7$ partial waves. Furthermore, our NNLO
result is globally similar to the one of the non-relativistic
approach. 

\section{Summary and perspective}
\label{Summary}
In this paper we have worked out the nucleon-nucleon interaction up to
NNLO in the framework of manifestly Lorentz-invariant ChEFT.  
We have renormalized the one-loop diagrams contributing to the TPE potential and the scattering amplitude by using the subtractive renormalization. 
In the large nucleon mass limit, the resulting TPE potential is consistent with its non-relativistic counterpart. 
Using the one-loop order approximation we calculated the NN phase
shifts and mixing angles for partial waves with the orbital angular momentum $l\geq
2$  
and compared the obtained results with the corresponding ones of the non-relativistic formulation. We found that the description of $D$ waves, particularly for $^3D_3$, 
is improved because of the relatively small contribution of TPE
diagrams. For the other peripheral partial waves, both approaches give
(globally) similar results.  For the $^3G_5$ partial wave,  
our results indicate a better convergence beyond NNLO compared to the non-relativistic approach.
      
Besides the higher partial waves, the description of the $S$ and $P$ partial waves and the deuteron bound state is most relevant in formulating the realistic NN force. 
Two strategies are available for considering the $S$ and $P$ partial waves at NNLO: 1) Restrict the non-perturbative treatment to non-singular LO potential of Ref.~\cite{Baru:2019ndr} and include the NLO and NNLO corrections perturbatively. 
This would allow to  systematically remove all divergences from the amplitude; 2) Treat the full NNLO potential non-perturbatively to obtain the NN scattering amplitude by solving the Kadyshevsky equation. 
The milder UV behavior of the effective potential and the scattering equation provide with a larger range of admissible cutoff-values, which is a welcome feature for the few/many-body calculations.\footnote{Both strategies are of course also applicable for peripheral phases.} 
Work along these lines is in progress.

\acknowledgements
This work was supported in part by BMBF (Grant No. 05P18PCFP1), by
DFG and NSFC through funds provided to the Sino-German CRC 110
``Symmetries and the Emergence of Structure in QCD'' (NSFC Grant
No. 11621131001, DFG Project-ID 196253076 - TRR 110), by DFG through the CRC 1044 ``The Low-Energy
Frontier of the Standard Model'' (Project ID 204404729 - SFB 1044), by the Cluster of
Excellence ``Precision Physics, Fundamental Interactions, and Structure of Matter'' (PRISMA$^+$, EXC 2118/1)
within the German Excellence Strategy (Project ID 39083149), by the Georgian Shota Rustaveli National Science Foundation (Grant No. FR17-354), 
by ERC AdG NuclearTheory (Grant No. 885150) and by the EU Horizon 2020 research and
innovation programme (STRONG-2020, grant agreement No. 824093).


\appendix


\end{document}